# How intellectual communities progress[1]

Lewis D. Ross
*London School of Economics*

**Abstract**

Recent work takes both philosophical and scientific progress to consist in acquiring factive epistemic states such as knowledge. However, much of this work leaves unclear what entity is the *subject* of these epistemic states. Furthermore, by focusing only on states like knowledge, we overlook progress in *intermediate* cases between ignorance and knowledge—for example, many now celebrated theories were initially so controversial that they were not known. This paper develops an improved framework for thinking about intellectual progress. Firstly, I argue that we should think of progress relative to the epistemic position of an intellectual community rather than individual inquirers. Secondly, I show how focusing on the extended process of inquiry (rather than the mere presence or absence of states like knowledge) provides a better evaluation of different types of progress. This includes progress through formulating worthwhile questions, acquiring new evidence, and increasing credence on the right answers to these questions. I close by considering the ramifications for philosophical progress, suggesting that my account supports rejecting the most negative views while allowing us to articulate different varieties of optimism and pessimism.

## 1. Introduction

Many of us care not only about the success of our own intellectual endeavours, but also those of the discipline to which we belong. Given this, it is discomfiting to note that some pessimism about progress within the philosophical community has gained traction lately (see van Inwagen 2004; Dietrich 2011; Chalmers 2015 for representative broadly pessimistic views and Stoljar 2017 for recent optimism).[2] In contrast, even though philosophers of science disagree on how to characterise it, there is a widely shared view both within and outwith academic circles that our scientific endeavours are progressing handsomely.

While there has been surprisingly little overlap between discussions of philosophical and scientific progress, a leading strategy in each debate takes acquiring *factive epistemic states* to be constitutive of progress. Recent work in philosophy of science put this commitment front and centre, simply arguing over which epistemic state best explicates progress—the epistemic view focuses on knowledge (see Bird 2007), the noetic view focuses on a broadly factive conception of understanding (see Dellsén 2016), and the semantic view focuses on truth or verisimilitude (see Rowbottom 2010 or Niiniluoto 2014). A similar focus on factive epistemic states is found within discussions of philosophical progress: for example, Stoljar (2017: 22) focuses on producing knowledge answering philosophical questions, and Chalmers (2015) looks for "large collective convergence" on true philosophical positions.[3] In what follows I work within this factive framework, even though it is controversial in various respects. For instance, while most parties to the present debate take understanding to be a broadly factive epistemic state, some question this assumption (e.g. see Elgin 2004; 2017). More radically, there is a rich tradition of thinking about progress in both philosophy and science in "anti-realist" terms, which on some views will entail rejecting explicating progress in terms of truth.[4,5] This paper will not attempt to refute non-factive approaches to progress, although much of what I say will be relevant to any theory in which the acquisition of *beliefs* figures in the conception of progress.[6]

---

[1] Forthcoming in Episteme, https://doi.org/10.1017/epi.2020.2
[2] Comments in Horwich (2012) also express sympathy with pessimism; see Slezak (2018) for recent criticism of Stoljar; Gutting (2016) argues to a more ambivalent conclusion, but nonetheless draws a contrast between philosophical and scientific progress.

A focus on acquiring epistemic states like knowledge is deeply woven into the literature on progress, but prevailing approaches leave two key questions unanswered. Firstly, we need to know what entity is the *subject* of the epistemic states purportedly constitutive of progress.[7] After all, epistemic states can be possessed by both individuals and by groups—so should we hold that progress is a matter of individuals or collectives gaining, e.g., knowledge? This question has ramifications for substantive verdicts about progress: the conditions for an individual versus a group possessing knowledge are by no means identical. And secondly, the focus on states such as knowledge leaves us unable to say much of substance about *intermediate cases*: cases in which an epistemic subject is on an upwards trajectory away from complete ignorance, and as such has made some partial progress, but has not yet fulfilled the demanding conditions required to possess knowledge. Such intermediate cases occur routinely. For example, they occur when some true theory has a modicum of support within an intellectual community but is subject to considerable disagreement. A lack of clarity about intermediate cases in the context of thinking about philosophical progress is particularly unfortunate—on one plausible view, an intermediate state between ignorance and knowledge is precisely where we find ourselves regarding many philosophical questions.

---

[3] Also see comments in Horwich (2012: 24) and Dietrich (2011: 335-6).

[4] For instance, this will encompass views that construe progress in terms of solving (or dissolving) problems. Within philosophy of science we might look to Kuhn (1962) or Laudan (1977) as examples. Within philosophy, certain conceptions of Wittgenstein's metaphilosophical project focus on dissolving philosophical puzzles through linguistic analysis.

[5] Some well-known objections to the factive approach—e.g. that many scientific theories are idealisations, or that progress occurs through adopting false theories—have been addressed by proponents of the factive approach (see Bird 2007a for an example).

[6] Another worry is that progress might consist in inventing certain instruments or methods, neither of which are laden with propositional content. As I see it, the invention of tools such as telescopes or techniques such as chromatography were progressive for epistemic reasons: viz. because we gained knowledge of how to use them, of the fact that using them might help answer certain theoretical questions, and because using them did in fact lead to more knowledge.

[7] For example, Dellsén's noetic theory focuses on individual scientists; Rowbottom is sympathetic to focusing on outputs like textbooks or lectures; while Bird looks to what scientists know collectively. These different starting-points are mostly assumed, yet each takes themselves to describe the same phenomenon.

This paper develops an improved framework for theorising about progress that resolves these issues. To preview, I argue that we should think of progress relative to the success or failure of *collective inquiry*. This framework has two components. After some necessary groundwork (§2), I firstly argue that our best theory of progress should focus on the factive epistemic states—such as knowledge—that enable an intellectual community to *settle inquiry qua group* (§3). Secondly, I use recent work on inquiry to suggest that collective inquiry typically progresses through a number of overlapping stages (§4). Thinking about these different stages of inquiry allows us to account for various *intermediate cases* of partial progress that traditional accounts are currently silent on, such as formulating worthwhile questions, acquiring new evidence, and increasing credence on the right answers to these questions. I close by considering the implications of this revised approach for angst about progress in philosophy, suggesting that focusing on collective inquiry enables rejecting outright despair, while providing resources with which to articulate different varieties of optimism and pessimism about philosophical progress.

## 2. Foundational Issues

We should start with some assumptions and desiderata to corral our discussion.

Firstly, judgements about progress are relative to some *end*. Often the relevant end is a self-consciously adopted *aim*: e.g. throughout 1944, the Manhattan project made considerable progress towards building an atomic bomb. But progress can instead be relative to an external *standard*: e.g. you might judge that your daughter is making good progress on her times-tables. Such a way of thinking, on which progress is end-relative, implies the value of progress depends on the value of the relevant end. This is right. After all, someone might lament the progress of vaccination-sceptics—such progress is lamentable because their endeavours are directed at a harmful end. We'll have more to say about this as we go on, but we are interested in the distinctively intellectual ends of academic inquiry.[8] Here, we will work within the dominant tradition supposing that the *ultimate* end for intellectual disciplines involves acquiring broadly factive epistemic state such as true belief, knowledge or understanding.

---

[8] Kitcher's (1993) discussion of the *impersonal epistemic* ends of science is instructive. On his view, the scientific discipline has aims which are impersonal insofar as they do not depend on the goals of any individual researcher, and epistemic insofar as they have an intellectual character rather than being directed at practical goals (e.g. advancing science for economic reasons).

Secondly, progress is relative to some *subject* being judged against an aim or standard. Different subjects can progress to different extents towards the same end; I aim to climb the mountain, the Joneses aim to climb the same mountain, and we may progress to different extents. As the introduction suggested, specifying the subject of progress is something about which there is not clear consensus. There are actually two distinct questions we must get clear on: (i) what exactly is the *explanandum* when theorising about progress? and (ii) having clarified the explanandum, what *epistemic subject* should we focus on when theorising about that explanandum? I'll chiefly be concerned with (ii), but first need to clarify (i). The explanandum is clearly the success or otherwise of some multi-agent endeavour. While focusing on cases in which an individual makes solitary progress and attempting to pinpoint what it consists in is a worthwhile project, this is not what recent work concerns. For one thing, within debates about philosophy, a dominant thought takes *absence of consensus* to be evidence for, or constitutive of, lack of progress. Moreover, the literature on scientific progress largely appeals to cases involving theories being disseminated among multiple agents. So our concern here should be with some sort of communal enterprise.

I propose that we focus on when the *intellectual community* engaged in the relevant academic inquiry makes progress. In what follows, I'll be talking rather loosely about *the* scientific and philosophical communities, but of course we may want to individuate intellectual communities with a finer grain (e.g. the community of cell biologists or metaphysicians). Focusing on intellectual communities is, I think, an informative presification of extant discussions about progress. But, even after clarifying that such multi-agent endeavours are the explanandum of recent debates, the puzzle regarding the relevant subject remains: one might explicate the progress of the relevant intellectual community by looking at the epistemic states possessed by its individual members or those it possesses *qua* group agent. In the next section I argue for the second strategy.

Before moving on though, we might briefly wonder—what binds intellectual communities together? We can begin by noting that they can be *partly* characterised with reference to the sorts of questions they are concerned with. Indeed, there is an idea that subject-matters can be identified with a question.[9] Hence, we have an attractive partial theory on which intellectual communities can be characterised by their commitment to certain sorts of inquiry. However, this is not a full characterisation—after all, two agents simply sharing a concern with the same question clearly does not implicate them in any collective endeavour. The issue of what exactly makes individuals participants in the sort of collective endeavour we are interested in is not one that I can settle here, but any focus on large intellectual communities will naturally tend towards supposing that the relevant commitment need not be explicit. One view in this vein is Bird's (2010a) discussion—drawing on Durkheim—of the *organic solidarity* that arises between academic practitioners in virtue of their mutual interdependence and division of labour. This approach is far from uncontroversial, however. For example, Wray (2007) offers scepticism about the idea that the different sub-disciplines in science add up to a unified and functionally integrated whole. Others, such as Gilbert (1989), might require more in the way of *explicit commitment* among group-members for them to be engaging in a collective inquiry. Accepting these views would push towards focusing on smaller intellectual communities than entire disciplines. Although I henceforth set aside this debate and focus on the scientific and philosophical communities in general, most of what I say will be equally applicable to a focus on smaller collectives.

---

[9] These questions can themselves be decomposed into sub-questions. One way to substantiate this approach is to appeal to partition semantics for interrogatives (e.g. see Cross and Roelofsen: 2.1.2 for an introduction) and take subject-matters to act as partitions on logical space. See Lewis (1998a; 1998b), Friedman (2013a), and Yablo (2014) for various discussions.

Having clarified that progress is relative to ends and to subjects, we need some desiderata for a satisfactory account. The first is that scientific and philosophical *regress* is a live possibility. This follows naturally from the fact that intellectual communities can become farther from achieving their aims. But regress is hardly mentioned in the literature. Therefore, one methodological innovation is using judgements about regress as an explanatory test for a good theory of progress.[10] Next, one difference in emphasis between debates about scientific and philosophical progress is the importance accorded to *novelty* and *convergence*. Philosophy of science shows a preoccupation with discovery. Conversely, regarding philosophy, the focus is on consensus on the correct views. A comprehensive theory of progress will do justice to the importance of both novelty and of convergence. And finally, progress seems to be a *gradable* notion—intellectual communities can make partial progress (and regress) relative to some intellectual end. More precisely, progress is gradable both with respect to the scope of novel contributions and to convergence. Some novel discoveries contribute more than others, and an intellectual community can converge on the truth to greater or lesser extents. Each of these thoughts should be accounted for in the framework we use to think about progress.

### 3. Progress and collective inquiry

---

[10] A straightforward way to incorporate such judgments is to assume symmetry between progress and regress—thus, if an intellectual community progresses towards achieving its epistemic ends in virtue of acquiring *x* (e.g. knowledge) it will regress in virtue of the loss of *x*.

I now argue that judgements about progress should be sensitive to the success or otherwise of *collective inquiry*, viz. whether or not relevant intellectual community *qua group* possesses the epistemic states needed to settle inquiry.

Focusing on *settling inquiry* will have three advantages. Firstly, it allow us to stay neutral on whether the epistemic, semantic or noetic account of progress is best—proponents of these views say that acquiring true belief, knowledge and understanding are the proper states with which to settle inquiry, and adjudicating between these theories is a delicate topic. Secondly, focusing on the settling of *inquiry* will neatly integrate points I make later on (in §4) about how moving through the different stages of inquiry is also progressive—it is progressive because it brings us closer towards settling inquiry. Thirdly and finally, focusing on settling inquiry will accommodate diverging views of a controversial type of case recently discussed by Alexander Bird and Jennifer Lackey. For convenience, I'll largely focus on settling inquiry by acquiring knowledge in what follows, but it should be clear how my discussion will encompass other factive epistemic states too.[11]

This claim that the best approach to progress will focus on *group* epistemic states is far from platitudinous. For, a natural supposition might be that the progress of an intellectual community is just directly relative to the progress of its members, without there being any need to posit additional epistemic states possessed by the collective. Indeed some traditional epistemologists might worry that talk of collective epistemic states does no important explanatory work. Here, I attempt to vindicate appealing to group epistemic states when it comes to theorising about progress.

Accepting the claim that our focus should be on group epistemic states doesn't involve thinking that what individual group-members know or produce is somehow unimportant. Rather, the suggestion is that we can do justice to the importance of success for individual academic inquirers (e.g. when a scientist acquires knowledge) by thinking about how this contributes to the success of collective inquiry for their intellectual community. However, notably, we will find that judgements about progress are not primarily sensitive to the presence of individual knowers. Taking the acquisition of group knowledge to be the relevant end to which progress is relative will allow us to appreciate why this is so.

---

[11] What about combining a *noetic* theory of progress with the focus on group epistemic states? Very little has been said about group understanding. However, I doubt that theorising about such a state has insuperable difficulties. 'Reductionist' accounts of understanding, on which understanding reduces to facts about what an agent knows or believes, have been defended in various places (e.g. Kelp 2017, Sliwa 2017; Ross 2018 for discussion). If understanding is reducible, there is no principled reason that group understanding could not similarly reduce to what the group knows and believes. Other dominant theories understanding in abilities/knowledge-how (e.g. see Hills 2017 or Elgin 2017). I lack space to discuss the details here but it seems equally possible for groups to possess practical knowledge.

Firstly, we can warm up with some instructive judgements. Let's start with a straightforward case: it wouldn't amount to scientific progress if a scientist, today, produced a new set of observational data establishing that mercury boils at around 356°C, and nor would pursuing such observational data be a suitable research programme for a professional scientist. Moreover, the failure of such research to be progressive doesn't change if there happened to be a very small number of scientists with false beliefs about the boiling-point of mercury. And finally, I suggest, it wouldn't amount to scientific *progress* if a colleague walked into the office of the ignorant scientists and informed them of this fact, thereby transmitting knowledge, and nor would it amount to scientific *regress* if a single scientist permanently forgot the boiling-point of mercury, thus losing their knowledge.

These simple cases suggest a broader lesson: our general judgements about progress cannot be easily explained *just* with reference to the epistemic position of individuals within an intellectual community. It is obviously a non-starter to suppose that *every* member of the group must have some belief for an intellectual community to progress. Nor does it require *most* members to have certain beliefs, as many academic advances are niche—they are only known to a small subset of the relevant community with the inclination and expertise to grasp them. For instance, in recent decades modern biology has afforded us with a much greater understanding of the highland midge; but very few scientists, even biologists, know much about midges. In this sense, progress is compatible with most of the relevant community lacking knowledge. It is also telling that judgements about progress don't necessarily march in lockstep with the *proportion* of group members with certain knowledge. If a biologist who studies termites expands her range and gets up to date on the latest work on midges, there doesn't seem to have been scientific progress just in virtue of her learning the pre-existing consensus among midge researchers. Nor does it seem to matter much, from the perspective of the community, if a research group acquiring knowledge of some new phenomenon contains $n$ or $n^{+1}$ members. Rather, it seems like the scientific community can definitively settle question notwithstanding the presence or absence of additional individual knowers.

All of these judgements are neatly explained by focusing on the presence and absence of *group knowledge*. For instance, regarding our initial simple case, the scientific community already knows the correct and complete answer to the question *'at what temperature does mercury boil?'*[12], explaining why further research into this question would be inappropriate and non-progressive. Moreover, as many social epistemologists accept, a group can know that *p* even if a minority of group members are ignorant about *p*; this explains the range of cases in which we do not judge there to have been progress (or regress) just if one individual learns (or forgets) that *p*, where this change is irrelevant to the overall epistemic standing of the group. Indeed, there is a compelling line of thought on which a group can know or believe something even without a majority its members possessing such knowledge. This view would have it that when there is division of labour among a group's members, a group can know that *p* just in virtue of those engaged in the relevant task knowing that *p*. The intuition behind this position is purely general, not just restricted to intellectual communities. For instance, we might suppose that the Police Dept. knows who committed the crime—indeed, we would say that they had *solved the crime*—even if only the detectives working on the case (rather than the hundreds of other employees) know whodunit. Groups of any size can designate particular members to inquire into questions on behalf of the wider group, and this provides a way for the group to acquire knowledge without necessitating that the content be widely believed by the rest of the group's members. This fact about social epistemology, conjoined with taking the subject of progress to be the intellectual community qua group, provides an explanation for why we do not require any particular *number* or *proportion* of a group's members to have certain beliefs or knowledge in order for the larger group to make progress.

---

[12] Some other questions admit of *partial* answers. For instance, the question *'who played football on Sunday?'* can admit distinct propositions as partial answers provided that more than one person played. It is plausible to suppose that knowing a more complete answer to a question is more progressive than knowing a less complete answer. Appealing to the way in which we can answer a question more or less completely might provide us with a way to model one axis of the gradability of progress mentioned earlier, namely progress with respect to the scope of novel contributions. See Pavese (2017) for a helpful critical discussion of the idea that different types of knowledge are gradable.

These data-points are all probative in favour of focusing on the epistemic position of the group subject when discussing progress. Further evidence is gained by considering scenarios involving a more fundamental disconnect between individual group members and the wider community. Here's a simple case, showing that changes in the epistemic position of individual members are not *sufficient* for progress.

> ***TRAGIC RESEARCH:*** Three logicians begin work on a research project, telling nobody else. The project comes to fruition and they prove some novel theorems. Before sending their work for review, they decide to take a hot-air balloon trip over Tuscany to celebrate their success. Tragically, the balloon explodes and no-one survives. Worse still, in thrall to a toxic work culture, each of the logicians had taken their laptops containing the research along with them. The novel research is destroyed—nobody ever learns anything about it.

This case is rather different from the aforementioned case of a Police Dept. solving a crime in virtue of just the detectives knowing whodunit. Here, it seems that the philosophical community has not made progress. Acquiring knowledge while *incidentally* being a member of some community is not sufficient for the community of which you are a member to make progress. Focusing on group epistemic states respects this judgement about progress—in ***Tragic Research***, the wider philosophical community lacks knowledge of the contents of the novel theorems the three logicians had proved.

Another case, inspired by a similar scenario discussed to a slightly different purpose by Alexander Bird[13], provides reason to think that individual knowledge might not even be *necessary* for progress:

> ***ARCHIVED RESEARCH:*** Professor Plum is a respected and conscientious researcher currently working on the biology of *Culicoides impunctatus*—the humble Highland Midge. As a topic on which there has been only a little scientific work, there are many rudimentary questions that remain unanswered. Plum is attempting to work out how far on average the male midge will range in search of food. She performs a series of studies that unequivocally support a certain answer: $n$ kilometres. The resulting paper is sent to a respected and well-indexed journal and published after the normal process of peer-review at $t_1$. A few years later, at $t_2$, Plum has died and everyone who read the paper when it appeared has either died or forgotten about it. Then, another biologist needs to find out about the behaviour of the male midge. They find Plum's paper after a quick search and quickly digest its contents, before citing it in his own paper at $t_3$.

Although ***Archived Research*** is a simplified case, its structural features are not uncommon. The massive distribution of labour in contemporary academic communities, along with the sheer volume of results being published, make it is natural to suppose that many findings might not currently be known by any particular researcher. So, what should we say about such cases?

Firstly, by stipulation, Plum made a novel discovery and made it accessible to the scientific community to be used in future research. Without yet taking a stance on the epistemic position of the group, I suggest that Plum's discovery constituted some degree of progress for the scientific community *at the time of publication*. The contentious question is: what happened at $t_2$? I suggest that whatever progress the community made *persisted* throughout the period of Plum's death until the work was taken up again at $t_3$. To bolster this thought, consider a competing framework looking only at individuals within the community. If progress were only sensitive to individual epistemic states, then we ought to judge there to have been *regress* when Plum died. However, this doesn't seem right—even though no individual believed the content at $t_2$, the wider community was in just as good a place as it was beforehand, still able to locate and use the research when needed. All of these data-points are captured by a focus on collective inquiry, just as long as we suppose that the community has successfully settled inquiry into the Q '*what is the range of the male midge?*' throughout the entire period.

---

[13] See Bird (2010a: 32).

Some theorists, notably Bird himself, controversially vindicate the epistemic improvement in the group's epistemic position by supposing that an intellectual community can retain *knowledge* of results that have been peer-reviewed and appropriately indexed, despite no longer being known or even believed by any individual group-member.[14] Others disagree. Lackey (2014) has argued against ascribing collective knowledge in such cases by drawing on plausible norms of assertion and action that knowledge is often thought to fulfil. For instance, suppose that the question being inquired into was not one of midge biology but rather concerned the safety of some drug. Would it be rational to prescribe the drug simply on the basis that the research vindicating its safety was in the archives although not actually believed or known by any individual group-member? We might suppose not. Rather, Lackey suggests that cases such as **Archived Research** *only put the group in a position to know*.

A focus on collective inquiry accommodates both Lackey and Bird's position. This is because there is a perfectly ordinary sense in which inquiry can be successfully completed just by being in a position to know something. For example, suppose that you are cooking an ambitious dinner for a pair of critical in-laws and want to have a reserve option in case things go awry. You decide you will need to find out the answer to the Q *'what is the phone-number of a very good Chinese restaurant?'* for emergency takeaway. You do some research, find out which restaurant has the most promising reviews, and you write down the phone-number on a slip of paper. Then, naturally enough, you forget the number—you no longer know what the number is. Nonetheless, I submit, you have successfully completed inquiry because you have put yourself in a position to readily know the answer to Q should you need to know it in the future. An intellectual community can collectively settle inquiry in the same way—by acquiring compelling sources of evidence that provide the answer to a question within the ambit of that community's intellectual aims, and enabling the answer to that question to feed back into future inquiry whenever it is needed. All of this is so, even without any individual member of the community possessing this evidence.

Let's take stock. I've argued that judgements about progress should be considered relative to the aim of an intellectual community settling inquiry by acquiring *group* epistemic states such as collective knowledge. Combining this with the earlier thought that subject-matters can be identified with sets of questions, we can end with a very rough gloss on what maximal progress looks like: when an intellectual community has settled on the complete answer to all of the questions constituting its distinctive subject-matter, then its work is done.

---

[14] For other potential sympathisers, see Rolin (2008) or de Ridder (2014).

# 4. Between ignorance and knowledge

Focusing on the aim of settling collective inquiry by acquiring states such as group knowledge is plausible at the upper echelons of intellectual progress, but is impoverished in a significant sense. Namely, looking only at states such as knowledge leaves us unable to say much about *intermediate cases* that intellectual communities invariably find themselves in. Intermediate cases occur when a group has intuitively made progress towards answering some question, yet lacks any factive epistemic state towards the answer. I'll firstly illustrate such cases, pinpointing why they are problematic for extant approaches to progress. Then, adapting insights from work on the nature of inquiry and the attitudes associated with it, I show how we improve upon extant accounts, vindicate the progressive nature of intermediate cases, all while still focusing on the success of collective inquiry.

## 4.1. Intermediate Cases

To make our discussion vivid, consider the following historical example:

> **CONTROVERSY.** At *t1,* the consensus view is that celestial objects other than earth are perfectly spherical crystalline bodies. Galileo then publishes *Siderius Nuncius* in 1610 detailing telescopic evidence suggesting that the lunar terrain is mountainous and crater-pocked. This evidence is deeply controversial at *t2,* splitting the intellectual community and prompting vigorous debate. Numerous intellectuals reject Galileo's claims, arguing that the apparent lunar imperfections were either projected there by his telescopic equipment, or that they were actually encased *within* a perfect exterior. Centuries later, at *t3,* Galileo's view has triumphed and the Aristotelean paradigm in astronomy is universally rejected.

Firstly, we can ask: does the intellectual community in question *believe* that the lunar terrain is mountainous and crater-pocked at *t2*?[15] Intuitively, it does not. The absence of group belief here is not only intuitive, but also follows from extant views on the nature of group belief. On summative views, a group belief is a function of individual belief (e.g. Quinton 1976)—for example, a group believes that *p* iff most of its members believe that *p*, or the group has privileged members (e.g. those delegated to inquire) who believe that *p*. Neither is the case here; by stipulation there is no majority for the Galilean view, and Galileo and his followers held no privileged position. On non-summativist views, groups can have beliefs that float free of what the majority, or the privileged members, believe; for instance, if the group has some decision procedure that yields a 'joint acceptance' of *p* among members (e.g. Gilbert 1989), or if we apply some aggregation function what individual members believe that yields an overall group judgement (e.g. List and Pettit 2011). Here, there was no such decision procedure, and no plausible function for judgement-aggregation would spit out a group belief in *p* when (at least) half of the members believe ~*p* and actively disbelieve *p*.

---

[15] Drake (1978) discusses the reception of *Siderius Nuncius* in the weeks and months following publication. I use a historical example only for illustration: if one has reservations about whether there really was a cohesive community in 1610, an analogous modern example works just as well.

Secondly, despite the dissensus, scientific progress had been made by the community at *t2*. As the publication of the *Siderius* is one of the most celebrated episodes of scientific history, disputing this would be deeply revisionary. And such revisionism is difficult to motivate; it doesn't seem plausible to wait until there is *outright* consensus on a theory before we judge that *some* progress has been made by the relevant community. For example, there are currently theories in theoretical physics that are subject of deep disagreement. While reaching full consensus on whichever of these theories turns out to be the most accurate would doubtless constitute progress, it is deeply counterintuitive to suggest we have made no progress whatsoever on answering the relevant questions. We can support the suggestion that progress is compatible with substantial disagreement by thinking about *regress*. Imagine that some correct scientific orthodoxy falls into controversy such that it can no longer be ascribed as a group belief. While it seems like *some* regress has been made, it doesn't seem right to say that there has been *complete* regress; that is, regress comparable to point at which the theory was not even under consideration by the relevant intellectual community.

These observations create a tension. We'd established that focusing on factive group epistemic states constituted a powerful framework for theorising about progress within an intellectual community, both tracking ongoing debates, and avoiding the pitfalls of focusing only on individual within the relevant community. However, if **Controversy** is a progressive episode in scientific history, it is not one that can be explicated by any of the following: group true belief, group knowledge or group understanding. For, it is platitudinous that: for a group to truly believe *p*, it must believe *p*. And, given minimal and eminently plausible assumptions: (i) for a group to know that *p*, it has to believe that *p*, and (ii) for a group to understand why *p*, it has to believe that *p* or at least believe some *q* that explains *p*.[16] In **Controversy**, at best, only some individuals within the group, rather than the group itself, possesses these epistemic states regarding Galilean astronomy.[17] This is the problem of intermediate cases: where an intellectual community has made progress without yet fulfilling the conditions to possess a group epistemic state like belief, knowledge or understanding.

---

[16] I lack space to comprehensively argue for these orthodoxies which are just one way to support the intuition that, in **Controversy**, the group doesn't collectively know, or understand the imperfect nature of the lunar surface. One might, of course, doubt these orthodoxies (e.g. by adopting a knowledge-first framework that denies the need to analyse knowledge in terms of belief (see, e.g. Bird 2010a); thinking that one can know on the basis of acceptance (for discussion see Wray 2001, 2007; Tebben 2019); or understand without belief (see Dellsén 2017)). However, this is no trouble for my argument here unless doubting these orthodoxies can yield a positive argument for attributing collective knowledge or understanding in **Controversy**.

[17] Here's an objection defused: intermediate cases can be explained by attributing some *different* group belief other than in the content *C* of the correct theory. For example, one might appeal to the belief/knowledge that <the evidence for *C* undermines the orthodoxy>, or <the evidence for *C* needs explained>, etc. However, this response does not capture all relevant cases. For, in the Galilean case,

Of course, one could say that cases such as **Controversy** are progressive in virtue of *promoting* or *approaching* whichever state one thinks is constitutive of progress. There is something right about this thought, but it is uninformative. After all, there may be a sense in which making coffee in the morning for the members of one's research group promotes or brings the group closer to the acquisition of novel knowledge (i.e. by waking them up), but it doesn't seem right to say that this act is progressive in and of itself.

In the next section I will clarify three genuinely progressive ways in which an intellectual community can come closer to settling inquiry. These three ways to advance towards settling collective inquiry, I suggest, are sufficient for intellectual progress in their own right yet are consistent with the idea that the ultimate aim is to acquire some factive epistemic state such as group knowledge.

### 4.2. Progress without belief

Simply looking at whether or not a group possesses knowledge turned out to be a blunt instrument for measuring progress, eliding incremental advances on the trajectory away from complete ignorance and towards knowledge. I propose that we can remedy this lack of nuance by looking at *other inquiry-related processes, states and attitudes* which are distinct from outright belief. In particular, we can extend pioneering work by Jane Friedman on individual inquiry in order to provide a richer framework for appreciating different types of collective progress.

Firstly though, we should pause to consider a natural thought: can we deal with cases such as **Controversy** simply by appealing to the familiar notion of *justification*? In this vein, one might suppose that progress occurred here in virtue of the group acquiring justification (whether in an outright or gradable sense) to believe in Galileo's astrology. My primary reservation with supposing that an appeal to justification can do all the work we need is that it will lack of the nuances of the account I will develop below. In this sense, the proof will be in the explanatory pudding only once I outline my account. Nonetheless, we can explicitly raise one worry now. Clearly, as the group fails to believe in Galilean astronomy, any appeal to justification to explain progress would have to concern propositional rather than doxastic justification. However, it seems that this focus is unable to capture the way in which *convergence* among group members of an intellectual community is a form of progress in its

---

the community was bitterly divided even in their *estimation* of the evidence: many didn't even believe that *C* should be taken seriously. If such cases are conceivable, then the progress cannot be explained by any belief or knowledge; if there's a split on whether research should be taken seriously, there isn't collective belief that it should be taken seriously.

own right. For, take a group that already has propositional justification to believe *p*. It seems progressive if this group becomes more confident in *p* over time, even if this happens without that group acquiring *more* justification for *p*. This intuition does justice to the preoccupation with convergence in the literature. But it cannot be captured by a view that just focuses on whether a group has or lacks propositional justification for some position; groups can become more or less confident (thus making progress) without gaining or losing propositional justification for the view in question. With this in mind, we can now turn to the more nuanced inquiry-based approach.

On Friedman's account of inquiry, developed in a series of papers, "a subject inquiring at *t* has an Interrogative Attitude at *t*."[18] (Note that Friedman uses 'attitude' as a loose catch-all term; nothing really hangs on it if—as I sometimes will below—you prefer to talk in terms of states or processes rather than attitudes). *Interrogative* attitudes are a range of different attitudes that we can hold toward some question, Q. These IAs include familiar folk-psychological states such as wondering, curiosity, contemplation, deliberation, suspension of belief, and so forth. The terminus of inquiry, on this framework, is something like a *settled belief* in a *complete answer* to Q. This explains why, for example, there seems to be some normative tension when we imagine an agent who knows that *p* (or even just takes themselves to have settled whether *p*) yet is also wondering or deliberating whether *p*. Settled belief is a distinct state to be contrasted with assigning a greater or lesser credence to some proposition. Compared with outright belief, a high credence is not incompatible with having an interrogative attitude. For instance, an agent can be pretty confident that *p* but still be deliberating whether *p*. Adopting these different interrogative attitudes—becoming curious about some phenomenon, wondering about a question, suspending belief until more evidence is acquired, contemplating the answer, deliberating over alternatives, and increasing one's credence on the way to a settled belief—are essential components of inquiring.

While Friedman's work mainly focuses on individual agents, it can naturally be extended to *collective* inquiry. For, it is natural to suppose that groups can be in a state of: inquiring into some Q, deliberating about answers to Q, and becoming increasingly confident in the answer to Q.[19] Certainly, it is a common enough to hear people ascribe these states to groups—it is perfectly felicitous to say (for example) that the scientific community is inquiring into some issue, or becoming more confident in some theory. I won't here get into questions about whether we should prefer a summative or non-summative interpretation of these ascriptions. Rather, my interest will be in showing that focusing on the different interrogative attitudes enables us to a better job of explicating progress than merely thinking about epistemic states like knowledge, justification, or outright belief.

---

[18] Friedman (*forthcoming*: 4). Also see Friedman (2013a; 2013b; 2017).
[19] It might be the case that certain of the states Friedman discusses, particularly curiosity, have a phenomenological component not easily realised by a group. As such, I won't be concerned with these states here.

Consider **_Controversy_** again. Instructively, it is precisely transitioning into and between different interrogative states that occurs in this case. The question we are judging progress relative to is: *'what is the lunar surface like?'* While there was no collective belief in a correct partial answer to that question, viz. that the lunar surface is imperfect and pock-marked, the intellectual community seemed to have progressed towards settling on the right answer just in virtue of considering Galileo's new evidence against the dominant incorrect view. Thus, vexingly for our promising framework for thinking about progress, the group acquiring knowledge was not necessary for partial progress. It is here that we can appeal to different stages of inquiry for a more nuanced account of how intellectual communities make progress.

The progress found in **_Controversy_** at *t2* compared to *t1,* so I argue, was at least threefold: (i) the intellectual community adopted an interrogative attitudes towards a significant question it had not yet correctly answered, (ii) the intellectual community acquired evidence supporting the answer to this question, and (iii) the intellectual community assigned higher credence to the correct answer to this question, an answer which equated to a fundamental fact about the nature of the cosmos. I suggest that each of these achievements—wondering about important new questions, acquiring evidence supporting the correct view, and becoming collectively confident about that view—are individually sufficient for an intellectual community to make partial progress towards answering the questions within its remit. These changes respect the importance of novelty and convergence in a comprehensive theory of progress. Adopting an interrogative attitude towards a question is a normal prerequisite for discovering *novel* truths, while increasing collective credence represents progress towards *consensus* within the intellectual community; acquiring evidence is important for both of these processes.

To further clarify the account, I'll now discuss three different interrogative attitudes roughly corresponding to three different stages of inquiry; transitioning into and through these stages of inquiry represents three different facets of group progress that do not require collective belief or any epistemic state entailing it. It may be that there are other ways in which an intellectual community can make partial progress but these, I suggest, are among the most important progressive elements of intermediate cases.

*Wondering:* A group moves into the state of wondering about Q by formulating a question and adopting some commitment to inquiring into Q. While wondering inevitably occurs alongside other interrogative attitudes, it typically begins prior to them—agents wonder about a question before (for example) deliberating over answers. Indeed, it is the fact that we are wondering about Q that usually explains why we are deliberating over answers to Q. Formulating and beginning to wonder about some significant Q to a state of wondering about Q constitutes partial progress towards answering Q.[20] One reason for this is a fact that we observed earlier— the ultimate aim of an intellectual community is to answer all of the questions properly within the ambit of their distinctive subject-matter, even when the group has not yet conceived of these questions. Identifying and then wondering about these questions is the first progressive stage on the trajectory to settling inquiry into these questions. Not only is identifying and formulating Q the first step towards answering Q, wondering about new and more fine-grained questions is an iterative process that is crucial for successful inquiry.[21] In particular, devising insightful research questions is particularly important for the acquisition of *novel* knowledge. For instance, once we affirmatively answered the question '*did Neanderthals control fire?*', it was further progress to then wonder *'whether Neanderthals created fire or merely harnessed naturally occurring sources of fire?'* A group, like our own scientific community, that was able to articulate this question, was doing better than the hypothetical group that was unable to form another relevant question. This is because wondering was the first step towards answering this question, beginning a process terminating in valuable new knowledge about our ancestors. Wondering about new questions is also important for answering *pre-existing* questions a group is already inquiring into: for instance, there are many cases where a group realises that, in order to answer some Q, it will be necessary to firstly answer Q*. For example, it might be necessary to know whether Neanderthals migrated to some area A (where evidence of purposive fire-making has been found) in order to know whether Neanderthals had the capacity to start fires. Until an intellectual community has *fully* answered all of the significant questions pertaining to its subject-matter, it does well by being in a state of ever more advanced wondering about them. Moreover, to press the point about comparative explanatory power raised earlier, note that wondering does not consist in the acquisition of propositional justification for some answer. Nor can it be explicated by appealing to the notion of evidence. Wondering about a new and important question is a distinct form of progress in its own right.

*Investigating and Evidence-gathering:* One interrogative process mentioned by Friedman is that of contemplation. A second, not mentioned by Friedman, but naturally taken to be a cognate, is investigation. One plausible way to think about contemplating Q is something like entertaining Q in thought, while investigating Q, roughly, can be thought of as seeking information relevant to answering Q. As I am conceiving of it here, contemplation can be a way of investigating a question, and is often how we investigate abstract philosophical questions. But many questions (including some philosophers seek answers to) demand purposive activity such as experimentation and building on prior research. However, *merely* investigating Q (whether via contemplation or otherwise) is not necessarily progressive; an individual or group might investigate some Q ineptly, learning nothing. Rather, investigation is important because (if things go well) it can yield *evidence* that facilitates answering Q. Although the exact conditions for evidence to come within the epistemic ken of a group has been little discussed it is nonetheless perfectly commonplace to suppose that (e.g.) the scientific community can acquire evidence. It isn't possible to settle knotty debates about the nature of evidence here, but we can provide a very rough gloss on the conditions for collective evidence-acquisition: an intellectual community acquires evidence when propositional content providing evidential support for some theory is made accessible to the relevant members of the community, such that they can use this content in reasoning and inference. This, paradigmatically, occurs when books and articles are published and disseminated within the relevant community. It is worth noting that this is also what occurs in the case of **Archived Research** discussed earlier—with the publication of his work, the community (rather than any individual) acquired evidence for Plum's answer to the midge question, and this evidence survived her death.[22] The collective acquisition of evidence that supports the correct answer to Q, I claim, is progressive. Taking evidence-acquisition to be a loci of progress coheres nicely with our focus on group epistemic states like collective knowledge as the terminus of inquiry; gathering evidence that supports a correct answer to Q is progressive because it facilitates the group *knowing* the answer to Q. This is true regardless of whether (e.g. like Lackey) one prefers a traditional approach or (e.g. like Bird) one prefers a knowledge-first account of collective knowledge. On a traditional approach, a group's evidence provides *justification* that enables their collective belief in an answer to Q to be knowledgeable; on a knowledge-first account, evidence is itself a form of knowledge that supports knowing the answer Q by entailing it. We celebrate historical episodes such as the publication of Galileo's *Siderius,* because Galileo promulgated powerful evidence for a ground-breaking intellectual discovery. Engaging in investigation that bears the fruit of evidence is one way for an intellectual community to progress, without acquiring a settled belief on the answer to whatever Q it is wondering about.

---

[20] This might need some qualification: it plausibly is not progressive to know the complete answer to Q and then start wondering about it all over again, although there may be interested cases involving undermining higher-order evidence that count against this thought.

[21] The extent of progress might be sensitive to the type of question being asked; some questions are plausibly more worthwhile than others. Indeed, some questions might not be worth answering at all.

*Deliberating and Credence-revision:* Another important part of answering a question is deliberating, viz. weighing the evidence for different answers on the way to settling on a belief. Groups like intellectual communities deliberate by directing the relevant group-members to consider the evidence for and against these views. Deliberating itself is not necessarily progressive; deliberation can tend towards the false just as well as the truth. However, when deliberating over answers to Q, groups become more or less confident in different positions.[23] One lesson we can draw from cases such as **_Controversy_** is that while something like consensus might be the ultimate aim for an intellectual community, it is more plausible to take the influence of agreement and disagreement on progress to come in degrees rather than as a binary enabler or disabler of progress (for instance, being determinative only when a community acquires or loses a collective belief). If the influence of agreement and disagreement on progress comes in degrees, it is not easy to capture this gradability by focusing only on binary, belief-entailing epistemic states that are not apt to be taken as degreed notions. Rather, this type of progress that falls short of consensus is better captured by appealing to the notion of a group's _credence_, viz. its degree of confidence in a given proposition. When a group becomes increasingly confident in the right answer, this constitutes progress; an essential part of an intellectual community transitioning from having an interrogative attitude towards Q towards settling on (and knowing) the correct answer is increasing the credence assigned to that answer. Progress in intermediate cases is not exhausted by increasing credence. This is because wondering about some significant Q and gathering evidence for the answer can be progressive even the group remains agnostic throughout this process. However, due to the fact that persistent disagreement is often a primary roadblock preventing a group transitioning from inquiring to having a settled belief, a group becoming more confident in a correct answer to Q represents one of the most important facets of intellectual progress. Appreciating the importance of credence-revision captures the sense in which, as we noted at the beginning of the paper, _convergence_ is a crucial component of progress for intellectual communities; groups make progress towards converging on the right answer by becoming more confident in that answer.[24] Focusing on collective credence also takes into the account the lesson of our earlier discussion; progress relative to consensus does not come all at once only when collective belief is gained—rather, it is sensitive to how confident the intellectual community is in the right answer.

---

[22] Of course, we must be careful to draw a distinction between cases like **Controversy** and cases like **Archived Research.** In both types of case there was evidence acquired by the group. However, in the latter case, the compelling nature of the evidence acquired by the group and the fact that it was not apt to elicit widespread disagreement meant that it was sufficient to *settle* inquiry into the relevant question. This is not the case in **Controversy**; this is an intermediate case of partial progress, because it is implausible to suppose that the relevant community had settled inquiry at *t2*.

The various interrogative attitudes and processes we have outlined do not involve group belief or knowledge. However, thinking about these other attitudes and processes affords us with a richer appreciation of the different types of progress a group makes on the way to settling a question. This allows us to explain why an intellectual community is doing well by transitioning to an in intermediate case between complete ignorance and knowledge: short of knowing the right answer, a community can make progress by formulating the right sorts of question, by gathering evidence, and by weighing up this evidence while becoming increasingly confident in the right answer.[25]

---

[23] I am supposing that a group will become more confident as its relevant members (respecting the ability of groups to delegate inquiry) become more confident. There is a small literature attempting to devise formal rules for aggregating group credences from individual beliefs and credences (see, e.g. Russell et. al. 2015; Dietrich forthcoming). This literature is complex and technical; I lack space to delve into it here.

[24] I am sympathetic to supposing that groups must increase their confidence in the correct view on the basis of _evidence_. Those who think of progress in terms of mere true belief might not require such a condition; this may be a mark against that view.

Finally, appealing to interrogative states doesn't only explain *progressive* intermediate cases, it also provides a helpful perspective on different scenarios involving *regress*. For example, focusing on collective credence allows us to explain why it is sufficient for partial regress if a group (at *t2*) becomes decreasingly confident in a true theory even if, prior to this at *t1*, the theory did not command quite enough support within the group to be attributed as a group belief. This is regress, even though no group knowledge has been lost, and the framework defended here explains why this is so: because the group has lost confidence in the right view. However, not all cases of regress can be given a credence-theoretic treatment; some require that we look to the absence of specific interrogative attitudes. For instance, suppose that a group has formulated an interesting question, started the process of investigating it, but all of the group's members remain avowedly neutral on the answer. It would seem to be regressive if (for, say, sociological reasons) this question became entirely neglected at some later time. But this won't be explicable due to any changing credence in the answer. Rather, regress has occurred because the group has stopped wondering about the question. Looking at interrogative attitudes not only explains why there can be progress without new belief, but also why there can be regress without the loss of belief.

Our discussion has suggested that progress can be made by acquiring a plurality of epistemic states and inquiry-related attitudes. Even granting that the *ultimate* goal for an intellectual community is to answer questions with some factive epistemic state—i.e. knowledge (on the epistemic view), understanding (on the noetic view) or true belief (on the semantic views)—I have argued that it is possible to make genuine intellectual progress short of acquiring any of these factive states.

This conclusion tacitly rejects what we might call *monism* about progress, a view on which all progress can be explained just in terms of a single epistemic state. On a monistic approach, any progress short of answering a question with a factive epistemic state would be explained by the acquisition of *other instances* of that epistemic state. So, for example, Bird defends the idea that we need *only* appeal to the acquisition of knowledge to give a full account of intellectual progress. Responding to the thought that we can progress towards knowing the answer to a question without acquiring knowledge, Bird suggests that:

> ... *the relevant developments that promote knowledge will themselves be knowledge*. For example, one may progress towards knowledge of whether some theory is correct by accumulating relevant evidence. If one accepts Timothy Williamson's equation of evidence and knowledge, then that evidence-gathering process will itself be the accumulation of knowledge. [Bird 2007: 83 emphasis added]

---

[25] A final question: is it progressive for a group to simply jettison, or become less confident, in a false belief? This is unclear. Rejecting false views is typically associated with acquiring evidence, and resuming inquiry into some question. My intuitions are muddy upon considering whether a group makes progress in virtue of replacing a false belief with no belief at all, without acquiring evidence, and without reopening inquiry.

The equation of evidence with knowledge therefore purports to provide an informative account of why intermediate cases are progressive, but without requiring that we go beyond looking for the acquisition of knowledge. One could go further in developing this monistic position by appealing to work concerning when *credences* constitute knowledge (e.g. see Moss 2013; 2016). This would provide further resources with which to theorise about intermediate cases while maintaining the view that only factive epistemic states are progressive.

Ultimately, I am sceptical that such monistic approaches to progress can succeed. There is no principled reason to accept that *questioning* is a form of knowledge, and there are well-theorised issues surrounding knowledge and disagreement which should make us doubt whether it is plausible to take the evidence acquired by intellectual communities in intermediate cases to constitute knowledge. And more broadly, the equation of evidence and knowledge faces various theoretical challenges.[26] Different issues would, of course, arise for a monist conception of progress centred on understanding or true belief. For instance, it is hard to see how increased credence could be given an understanding-theoretic treatment. But I leave these challenges for others to take up. My own view has it that progress can be realised by a plurality of states and attitudes, each of which can be seen as advancing towards the ultimate end of answering a question.

## 5. Conclusion & Coda on Philosophical Progress

This paper argued that the best framework for measuring progress in academic disciplines should focus on the epistemic standing of intellectual communities *qua group*, rather than on their individual members. However, reflecting on celebrated examples of intellectual progress revealed that the dominant approach—taking epistemic states such as knowledge or belief to be exhaustive of progress—failed to do justice to ubiquitous *intermediate cases* of partial progress, where an intellectual community is on the trajectory away from ignorance and towards knowledge. A more nuanced account of progress, I suggested, is facilitated by paying attention to less heavily theorised attitudes and processes associated with inquiry. Using this approach, we were able to capture some different varieties of progress that intellectual communities can make: formulating significant research questions; gathering evidence; and becoming collectively more confident in the correct answer, none of which necessarily involve the acquisition of collective belief or knowledge.

---

[26] E.g. see Brown (2018) for recent critique.

I would like to close by considering how this general framework impinges upon our thinking about philosophical progress. A number of philosophers have endorsed pessimistic evaluations of philosophical progress, derived from an apparent lack of consensus on the big philosophical questions within the philosophical community. One upshot of focusing on progressive attitudes that fall short of belief is that it enables us to reject outright pessimism about philosophical progress, but without committing to the strategy (e.g. taken by Stoljar 2017) of claiming that the philosophical community has actually answered many fundamental questions. Instead, we can press the thought the philosophical progress consists in formulating the right sorts of questions, in gathering evidence for correct theories, and in the increasing popularity of correct theories (if there are such in the corpus) within the philosophical community. Each of these forms of progress is consistent with the absence of collective belief in particular theories. Of course, it is not entirely transparent to us whether we have identified the correct theories, and hence, the extent of our progress is not entirely transparent. But this lack of transparency is an entirely typical feature of progress; if a group is traversing a ridge in foggy conditions, it may be unclear to them—despite their best efforts!—whether or not they are moving in the right direction.

In a broader sense, thinking about different stages of inquiry allows us to provide fine-grained loci for different varieties of optimism and pessimism about philosophical progress. For instance, one might endorse optimism about the philosophical community's ability to ask important questions and gather evidence for theories, but pessimism about our ability to collectively settle on particular views. A valuable project for the future will be to try and understand what explains these strengths and weaknesses. One important way to approach this project will be to look at other intellectual communities as comparators; looking at how they transition through the different stages of inquiry, how they gather evidence, and how they use this evidence to collectively settle on theories. The framework outlined in this paper has provided a way to approach these comparisons—this, I hope, constitutes progress of a sort.[27]

---

[27] Ironically for a paper on progress, my own progress in writing this article was particularly painstaking. Many people offered advice and support, in particular two anonymous referees for this journal, Tuuli Ahlholm, Jessica Brown, Matt McGrath, Miguel Egler, Hannah Rose Blakeley, Brian Weatherson, my friends and colleagues in the Arché Research Centre, and audiences at the Social Epistemology Network conference in Oslo and the St Andrews 'Friday Seminar'.